\documentclass[aps,pra,twocolumn,noshowpacs]{revtex4}
\usepackage{graphicx,epsfig, subfigure}
\usepackage{amsbsy}
\usepackage{amssymb}
\usepackage{amsmath}
\usepackage{hyperref}

\begin{document}

\newcommand{\eqnref}[1]{Eq.~\ref{#1}}
\newcommand{\figref}[2][]{Fig.~\ref{#2}#1}
\newcommand{\citeref}[1]{Ref.~\cite{#1}}

\newcommand{\Sperp}{S}
\newcommand{\Q}{Q}

\title{Dynamic stabilization of a quantum many-body spin system}
\author{T.M. Hoang, C.S. Gerving, B.J. Land, M. Anquez, C.D. Hamley,  and M.S. Chapman}

\affiliation{School of Physics, Georgia Institute of Technology,
  Atlanta, GA 30332-0430}

\maketitle

{\bf It is well-known that unstable equilibria of physical systems can be dynamically stabilized by external periodic forcing \cite{Landau76}. The inverted pendulum stabilized by vibrating the pivot point (`Kapitza's pendulum') provides a classic example of this non-intuitive phenomenon and was first demonstrated over 100 years ago \cite{Stepheson08,Kapitza51}. Dynamical stabilization has a broad range of applications including rf Paul traps and mass spectrometers \cite{Paul90}, particle synchrotrons \cite{Courant52}, and optical resonators \cite{Siegman}.
Here, we demonstrate dynamic stabilization in the collective states of an unstable strongly interacting quantum many-body system by periodic manipulation of the phase of the states.
The experiment employs a spin-1 atomic Bose condensate that has spin dynamics similar to the Bose-Hubbard two-well system and is analogous to a non-rigid pendulum in the mean-field limit \cite{Smerzi97,Zhang05}. The condensate spin is initialized to an unstable (hyperbolic) fixed point of the phase space, where subsequent free evolution gives rise to spin-nematic squeezing \cite{Duan02,Mustec02,Hamley12} and quantum spin mixing \cite{Law98,Gerving12}. To stabilize the system, periodic microwave pulses are applied that manipulate the spin-nematic
quantum correlations
and coherently limit their growth.  The range of pulse periods and phase shifts with which the condensate can be stabilized is measured and compares well with a linear stability analysis of the problem.
These experiments demonstrate new methods of manipulating out-of-equilibrium quantum many-body systems, drawing together ideas from classical Hamiltonian dynamics and quantum squeezing of collective states.
}
\begin{figure*}[t!]
	\begin{center}
	\includegraphics[width=6in]{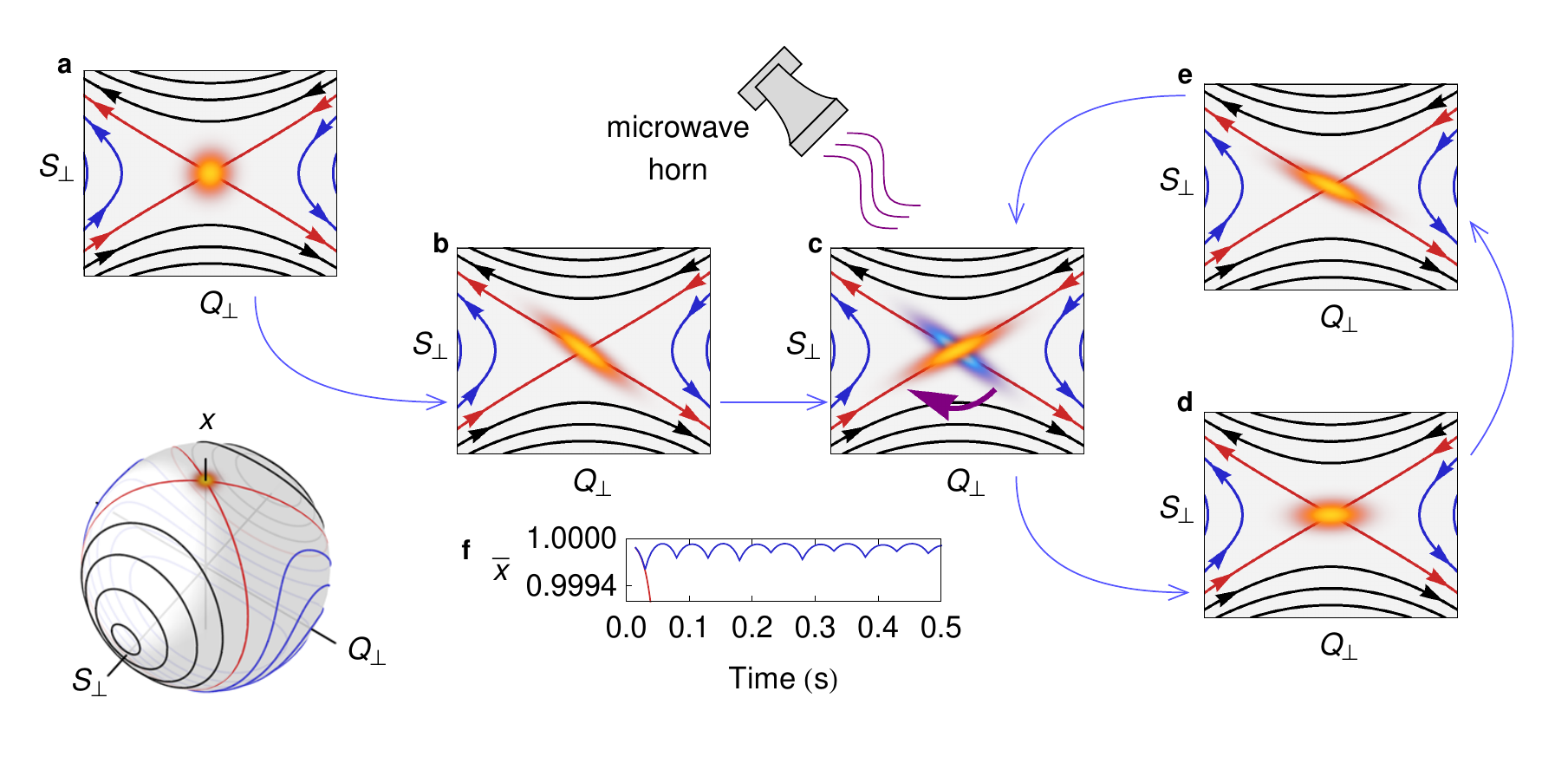}
	\end{center}
\caption{\textbf{Illustration of the dynamic stabilization method.} (a) The condensate is initialized at the pole of the spin-nematic sphere, $x=1, \Sperp_\bot = \Q_\bot = 0$.  The condensate has Heisenberg-limited uncertainties in $\Sperp_\bot$ and $\Q_\bot$. (b) Initial free evolution of the condensate produces spin-nematic squeezing along the diverging manifold of the separatrix.  (c)
The quantum state of the condensate is quickly rotated to the converging  manifold of the separatrix using a microwave field pulse.
(d) Subsequent free evolution unsqueezes the condensate, returning it close to the original state. (e) Continued free evolution again generates spin-nematic squeezing.  (f) Long term stabilization is achieved by repeating the (c,d,e) sequence (blue line) whereas the unstabilized condensate rapidly evolves away (red line).}
\label{fig:Concept}
\end{figure*}

Recent advances in ultracold atomic physics provide opportunities to investigate unstable equilibrium phenomena of interacting quantum many-body systems featuring well-characterized and controllable Hamiltonians \cite{Bloch08}. By changing the dimensionality of the system, tuning the interaction strength \cite{Chin10}, or magnetically quenching a spin system \cite{SK12}, it is possible to study excitations across a quantum phase transition described by generalizations of the Kibble-Zurek mechanism and apparent relaxation to non-thermal steady states in a well-isolated quantum system \cite{Polkovnikov11}. Beyond these fundamental issues, non-equilibrium dynamics can generate squeezed states and non-Gaussian states that are potential resources for quantum enhanced measurements \cite{Ma11} and quantum information processing \cite{Braunstein05}. Dynamic stabilization of non-equilibrium many-body Bose-Einstein condensates (BEC) has been suggested by tuning the sign of the scalar \cite{Saito03,Saito07,Compton12} and spin-dependent \cite{Zhang10} interaction strength and by time-varying the trapping potential in a double well BEC \cite{PhysRevA.80.053613,Boukobza10,Sols04}. Related ideas have been employed to suppress tunneling in optical lattice systems as a means to control the superfluid-Mott insulator phase transition \cite{Zenesini09}.

Here, we demonstrate dynamic stabilization of the non-equilibrium dynamics of a multi-component spinor condensate. The spin dynamics following a magnetic quench are well-described by an unstable inverted pendulum in the spin-nematic degrees of freedom \cite{Gerving12}. The non-equilibrium dynamics are stabilized by periodic application of phase shifts to the collective states of the system.
The experiment is conceptually related to spin decoupling or refocusing techniques used in nuclear magnetic resonance (NMR) \cite{Ernst} and bang-bang control of non-interacting two-level quantum systems (qubits) in quantum information processing \cite{Viola98}, although here it is applied to the collective dynamics of an interacting quantum spin system.
In future work, it should be possible to engineer highly entangled states of the system (including Schr\"{o}dinger cat-like states) using extensions of these ideas \cite{Leung12} and to demonstrate other fundamental phenomena such as Berry phases \cite{Li11}.

\begin{figure*}[t!]
	\begin{center}
	\begin{minipage}{7in}
		\begin{minipage}{3.4in}
		\hspace{1in}
		\end{minipage}
		\begin{minipage}{3.4in}
		\includegraphics[scale=1]{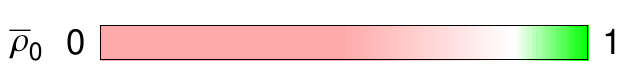}\\						
		\end{minipage}
	\end{minipage}
	
	\begin{minipage}{7.0in}	
		\begin{minipage}{3.45in}
		 	\includegraphics[scale=1]{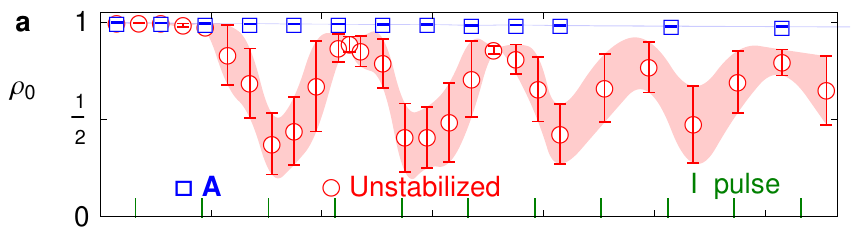}\\	
		 	\includegraphics[scale=1]{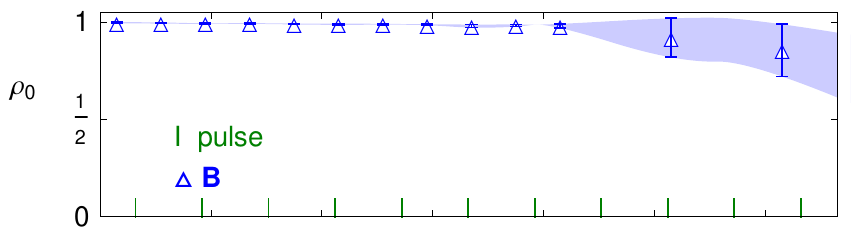}\\	
		 	\includegraphics[scale=1]{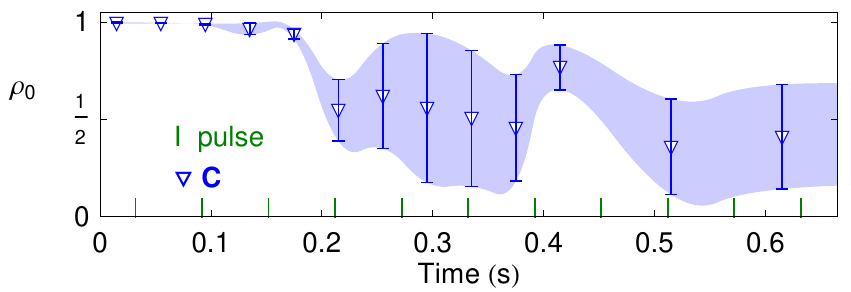}		 		 		
		\end{minipage}
		\begin{minipage}{3.45in}
			\includegraphics[scale=1]{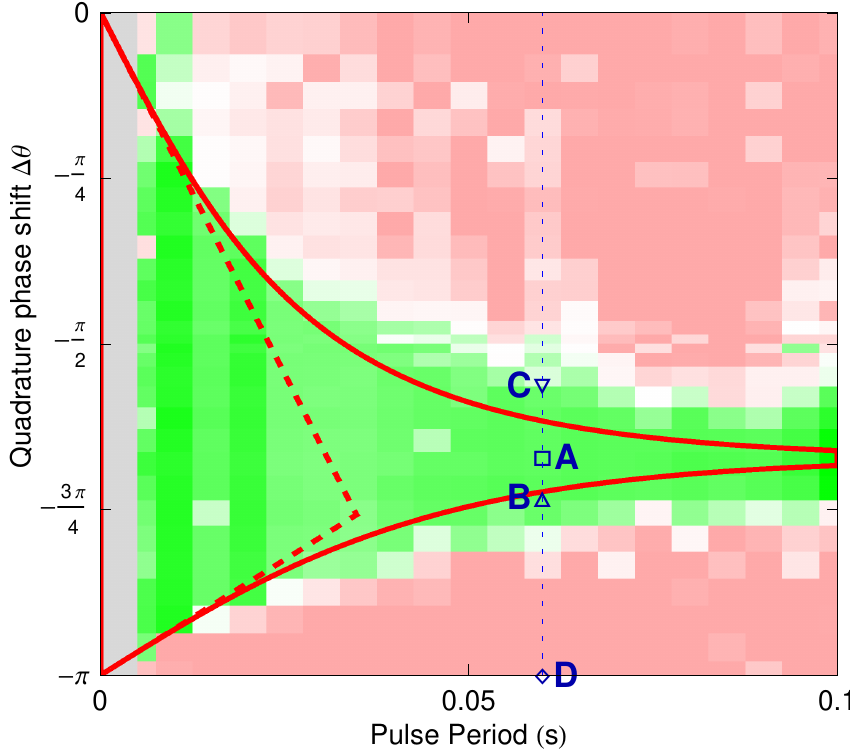} 			
		\end{minipage}	
		\begin{minipage}{7.0in}
		 	\includegraphics[scale=1]{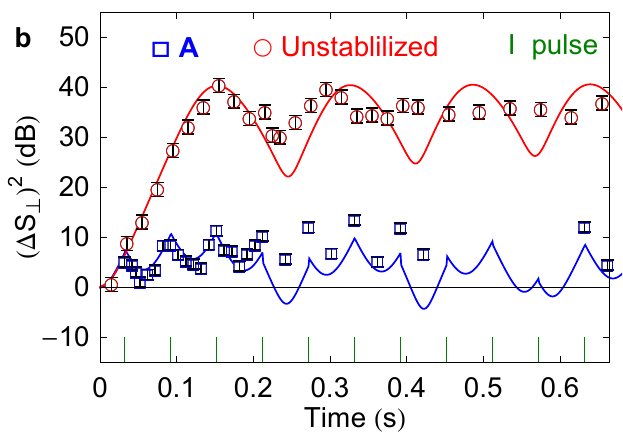}				
		 	\includegraphics[scale=1]{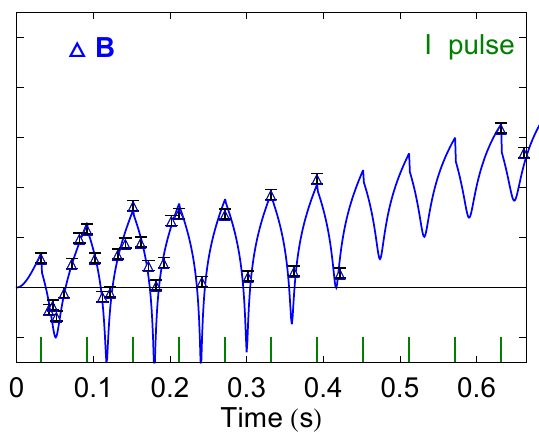}				
		 	\includegraphics[scale=1]{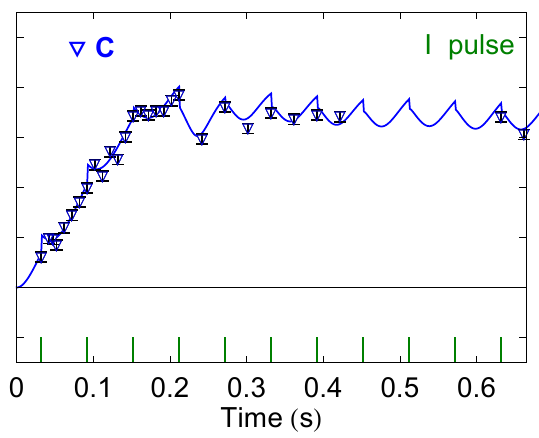}				 		 	
		\end{minipage}			
	\end{minipage}

	\end{center}
    \caption{\textbf{Stabilized dynamics and stability mapping}. The labeled cases correspond to period and quadrature phase shifts ($\tau, \Delta \theta$) on the stability mapping.  The period is $60~\mathrm{ms}$ for all while the phases shifts are Case A $\Delta \theta=-0.65 \pi$, Case B $\Delta \theta=-0.72 \pi$, Case C $\Delta \theta=-0.56 \pi$, and Case D $\Delta \theta=-\pi$. Pulse timings are shown as green ticks.  (\textbf{a}) Ideal stabilized population $\rho_0$ dynamics (top) Case A(blue square) versus unstabilized (red circle), (middle) dynamics near the stability edge where the dynamics eventually destabilize Case B(blue up triangle), and (bottom) dynamics outside the stability region Case C(blue down triangle).   
    (\textbf{b}) Variance of the transverse magnetization $\left(\Delta S_\bot\right)^2$ for (left) Case A versus unstabilized (red circle), (middle) Case B, and (right) Case C. Solid lines are quantum simulation. (\textbf{c}) Map of the experimental stability region (green) shown with the analytic stability solution (red solid line) for $\rho_0$ population after 185 ms of evolution. Also shown is the `robust' region where the mean effective $q$ is stable (red dashed line).}
\label{fig:StabilityMapuWave}
\end{figure*}

\section{Experimental concept}
The experiment uses small spin-1 rubidium-87 condensates containing just a single domain, such that the dynamic evolution occurs only in the internal spin degrees of freedom. This simplifies the  many-body problem to a zero-dimensional system (a `microcondensate' \cite{Lamacraft11}) of interacting quantum spins that permits description using  macroscopic quantum variables such as the collective spin operators \cite{Law98,Duan02}, $\hat{S}_i = \sum_{j,k} \left(\mathbf{S}_i\right)_{jk} \hat{a}_j^\dag \hat{a}_k$ ($i \in x,y,z$ and $j,k \in 0,\pm1$), where $\hat{a}_j^\dag$ is the raising operator for the $j^{th}$ $m_f$ state of the condensate. 
In this case, the many-body Hamiltonian up to constants can be written \cite{Hamley12}
\begin{equation}
	\hat{\mathcal{H}} =   \lambda \hat{S}^2 +\frac{q}{2}\hat{Q}_{zz} \nonumber
\label{Hamiltonian}
\end{equation}
where $\lambda \propto N^{-3/5}$ is the spinor interaction energy integrated over the condensate, $\hat{S}^2=\hat{S}_x^2+\hat{S}_y^2+\hat{S}_z^2$ is the total spin operator, $q \propto B^2$ is the quadratic Zeeman energy, $\hat{Q}_{zz} =\frac{2}{3} \hat{N}_1+\frac{2}{3}\hat{N}_{-1}-\frac{4}{3}\hat{N}_0$ is an element of the spin-1 nematic (quadrupole) tensor, and $\hat{N}_k=\hat{a}_k^\dag \hat{a}_k$, ($k=0,\pm1$) is the number operator for each of spin projections of the condensate.  The Hamiltonian conserves the total number of atoms $\hat{N}=\hat{N}_1+\hat{N}_{-1}+\hat{N}_0$ and the magnetization $\hat{S}_z=\hat{N}_1-\hat{N}_{-1}$, and exhibits a quantum phase transition at $q=-4N\lambda$ for $\lambda<0$ as is the case here.
For $S_z=0$, the condensate has an unstable equilibrium point at $N_0=N$ for small quadratic Zeeman energies; in the neighborhood of this point, the linearized equations of motion in the rotating frame are given by
\begin{eqnarray}
	\dot{\hat{\Sperp}}_{x} &=&- q \hat{\Q}_{yz} \nonumber\\
	\dot{\hat{\Q}}_{yz} &=&  \left( 4N\lambda + q \right) \hat{\Sperp}_x
\label{sho}
\end{eqnarray}
with identical equations for ($\hat{S}_y,\hat{Q}_{xz}$) reflecting the rotational symmetry about the $z$ axis (Supplemental Information). These equations describe a quantum (many-body) inverted harmonic oscillator that is inherently unstable.

The nature of the instability is illustrated with aid of the mean-field spin-nematic phase space of the  Cartesian components of the spin vector $S_i$ and nematic (quadrupole) tensor $Q_{ij}$ \cite{Hamley12,Plimak2006311}. In this space, the mean-field spin dynamics are represented on a unit sphere with axes  $ \{\Sperp_\bot ,  \Q_{\bot } , x \}$ where $\Sperp_\bot^2 = \langle \mathbf{S}_x\rangle^2+\langle \mathbf{S}_y\rangle^2 $, $\Q_\bot^2 = \langle \mathbf{Q}_{xz}\rangle^2 + \langle \mathbf{Q}_{yz}\rangle^2$, and $x = 2N_0/N-1$ (Supplementary Information). This sphere is shown in \figref[a]{fig:Concept} together with the mean-field dynamical orbits of the system for $q<4N|\lambda|$. The unstable fixed point is located at the pole of the sphere at the intersection of the two manifolds of the separatrix that divide the space into phase-winding and oscillatory phase orbits. The mean-field phase space is functionally identical to the symmetric double-well Bose-Hubbard model \cite{PhysRevA.82.053617}, and both can be described using a classical non-rigid pendulum \cite{Smerzi97,Zhang05} where the unstable fixed point corresponds to an inverted pendulum.

\begin{figure}[t!]
	\begin{minipage}{3.4in}
	 		\includegraphics{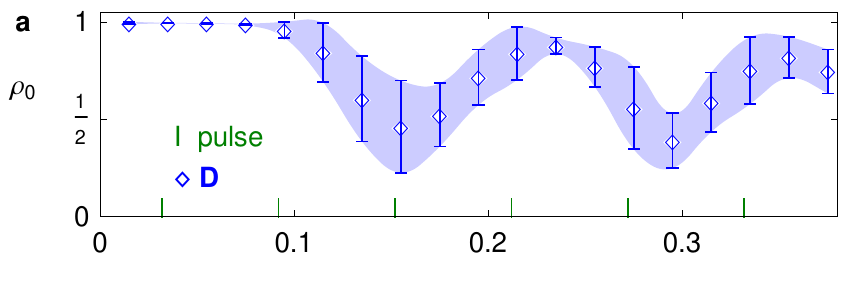} 		
	\end{minipage}	 	
	\begin{minipage}{3.4in}
	 		\includegraphics{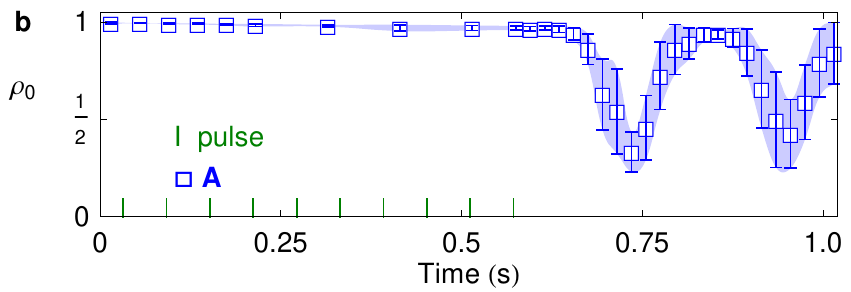}	 		
	\end{minipage}	 	
    \caption{\textbf{Maintenance of spin-mixing.}  (a) Stabilization dynamics for a quadrature phase shift of $-\pi$ pulse (Case D in \figref{fig:StabilityMapuWave}). (b) Stabilization dynamics with most stable parameters (A) for 572 ms followed by free evolution.}
\label{fig:StabilizedSpinMixing}
\end{figure}

The controllable quantum phase transition can be exploited to initialize the condensate in a quantum limited uncertainty centered on the unstable point \cite{Hamley12}.
In the mean-field limit, this is a non-evolving state, however the full quantum solution shows evolution that generates Gaussian squeezed states at early times (localized near the pole) and a rich variety of non-Gaussian states at later times as the system evolves along the separatrix \cite{Gerving12}. The quantum solution only converges to the mean-field logarithmically with the number of particles \cite{PhysRevA.82.053617,PhysRevLett.86.568} 
making it necessary to take quantum fluctuations of the initial state into account.  Quantum dynamics can be recovered semi-classically by using the classical equations of motion with a quasi-probability distribution such as shown in \figref{fig:Concept}.
In the classical phase space initial evolution in the neighborhood of the hyperbolic fixed point reduces the fluctuations along the converging manifold of the separatrix and grows the fluctuations along the diverging manifold of the separatrix; together, these dynamics create squeezing in the spin-nematic phase space through the growth of quantum correlations.
The continued growth of the quantum correlations eventually destabilize the system and lead to evolution away from the fixed point \cite{Gerving12}.

Dynamic stabilization can be achieved by preventing the build-up of these correlations.  In our experiment, this is accomplished using periodic phase shifts of the spinor wavefunction that manifest as a rotation in the $\Sperp_\bot , \Q_\bot $ plane about the origin.  The concept is illustrated in \figref[a-d]{fig:Concept}.  During the initial evolution to a squeezed state, the squeezed uncertainty ellipse aligns along the diverging manifold of the separatrix. In order to prevent further growth of the correlations, the state is quickly rotated to the converging separatrix using a microwave pulse (Methods).
Further evolution `unsqueezes' the condensate and returns it approximately to the initial state, thereby stabilizing the system.  Long-term stabilization is realized by periodic repetition of this cycle. Although we discuss the technique using a rotation angle corresponding to the angles between the manifolds of the separatrix, the condensate can be stabilized for a range of angles as will be shown.

\section{Observation of dynamic stabilization}

The experimental results demonstrating dynamic stabilization of the condensate are shown in \figref{fig:StabilityMapuWave}. The time evolution of the spin population $\rho_0=N_0/N$ is shown in \figref[a]{fig:StabilityMapuWave} for different microwave pulse parameters chosen to produce a stabilized condition (Case A), a marginally unstable condition (Case B), and a more unstable condition (Case C). The unstabilized dynamics showing free evolution spin-mixing is shown for comparison.  In the stabilized cases, the pulse period is 60~ms with the first pulse at 32~ms after the quench.  The difference between Cases A, B, and C is the size of quadrature phase shift applied per pulse.  Each measurement is repeated 10--15 times and the mean and standard deviation are shown with the marker and error bar.  The region encompassed by the standard deviation is indicated by shading to guide the eye and indicate the growth of the fluctuations.
We have performed two additional measurements shown in \figref{fig:StabilizedSpinMixing} to verify that stabilizing pulse maintains the quantum features of the spin dynamics. In the first, we have studied the evolution of the condensate under periodic pulses with $\Delta \theta = -\pi$ (Case D) which should have no effect on the dynamics.  Here we have verified that the condensate undergoes normal quantum spin mixing on the same time scale as without stabilization pulses (\figref[a]{fig:StabilizedSpinMixing}) \cite{Gerving12}.  In the second, we have turned the stabilization pulses off after 572~ms and verified that the system again undergoes normal spin-mixing (\figref[b]{fig:StabilizedSpinMixing}).  The fact that these two experiments demonstrate spin-mixing on the same time scale as the unstabilized case is important in that spin-mixing from the $m_f=0$ state is sensitive to any noise seeding the initial states \cite{Klempt10,Gerving12}.  Extra noise is particularly noticeable in length of the initial pause (or `break time')  in the dynamics of $\rho_0$ where any added noise decreases its length.  Even after more than half a second of stabilization, this pause is still 100~ms in length.

Measurement of the spin population $\rho_0=N_0/N$ corresponds to a measurement of the projection of the spin-nematic sphere on the polar axis. Hence this metric is admittedly less sensitive to early dynamics of the state initialized at the pole and does not directly reveal the growth and control of the quantum fluctuations of the initial state discussed in the introduction. To access this physics more directly, we have measured the evolution of the transverse magnetization $(\Delta \Sperp_\bot)^2$ by performing an RF rotation of the spin-1 state (Methods). These measurements are shown in \figref[b]{fig:StabilityMapuWave} for the same stabilization pulse parameters as above. Each measurement is repeated 30 times in order to accurately determine the variance.  
These results are compared with a fully quantum calculation where the initial state is a Fock state with 45,000 atoms in $m_f=0$ and the atom loss is accounted for by time varying the spinor dynamical rate detailed in \citeref{Gerving12}.
For the unstabilized condensate, the fluctuations grow exponentially by a factor of $10^4$ within 150~ms and eventually execute small oscillations near the maximum value.  When the condensate is stabilized, the fluctuations of  $\Sperp_\bot$ exhibit periodic growth and reduction during each pulse cycle, which reflect the squeezing and unsqueezing of the condensate. In the short timescale up to 0.4~s, the data is in good agreement with the quantum calculation.  The stabilized data (Case A) show the expected periodic evolution of the fluctuations and also show a dramatic reduction of the fluctuations compared with the unstabilized condensate.  For Case B, the fluctuations show squeezing  below the standard quantum limit (SQL) indicated by the 0~dB line, while in Case C, the fluctuations grow similarly to the unstabilized case.

We have also investigated the range of pulse periods and quadrature phase shifts that provide stabilization of the spin dynamics.  These measurements are shown in \figref[c]{fig:StabilityMapuWave}, which displays a map of the stability region versus pulse period and quadrature phase shift (modulo $\pi$, the periodicity of the phase space).  The stability criteria applied is $\bar{\rho}_0 > 0.85$ for 3 runs at 185~ms of evolution indicated by the green region.  The locations corresponding to the time sequences (Cases A-D) are indicated on the stability map. The data is compared with a linear stability analysis similar to methods used for optical resonators (Supplemental Information). This stability condition from this analysis is shown as the solid red lines in \figref[c]{fig:StabilityMapuWave} for the measured spinor dynamical energy $c\equiv 2N\lambda =- 2\pi\hbar \times 7.2(2)$~Hz and the measured magnetic field $B = 220(10)$~mG that determines the quadratic Zeeman effect $q = 2\pi\hbar \times 71.6 \times B^2~\mathrm{Hz/G^2}$.  
The data is in good overall agreement with the theory: for shorter pulse periods, the condensate is stabilized with a wide range of quadrature phase shifts, while for long pulse periods, the range of quadrature phase shifts capable of stabilizing the dynamics shrinks and reaches an asymptotic value close to the angle between branches of the separatrix, $\Delta \theta = \cos^{-1}(-1-\frac{q}{c})$.
The green region extends slightly beyond the linear stability analysis because for the marginally unstable cases, the dynamics have not had enough time to `fall off' the top of the sphere in 185~ms.

\section{Discussion}

The experiments presented above demonstrate dynamical stabilization of the spin dynamics of a spin-1 condensate. This is a many-body effect in that the spin dynamics of the condensate are driven by coherent collisional interactions in a 45,000 atom Bose condensate. The experiments reveal genuine quantum dynamics beyond the mean field as demonstrated by the control of the quantum fluctuations $(\Delta \Sperp_\bot)^2$. We stress, however, that the stabilization technique is applicable for states with classical noise or quantum noise. This same point is valid for an inverted simple pendulum: both a classical and quantum pendulum can be stabilized in the inverted configuration using the same protocol. Indeed, our experiment provides a compelling illustration of the quantum-`classical' correspondence principle where `classical' in this case refers to the mean-field description of the condensate wavefunction. For both our system and the inverted simple pendulum,  Heisenberg-limited states as well as states with excess noise can be stabilized using the same techniques. In this sense, the claim for `quantum' control of the dynamics (beyond the trivial point that the condensate is an inherently quantum entity) rests on the fact that the measured quantum fluctuations and characteristic features of quantum spin mixing are preserved by the method.

\begin{figure}
	\begin{minipage}{3.4in}
	 		\includegraphics[width=3.4in]{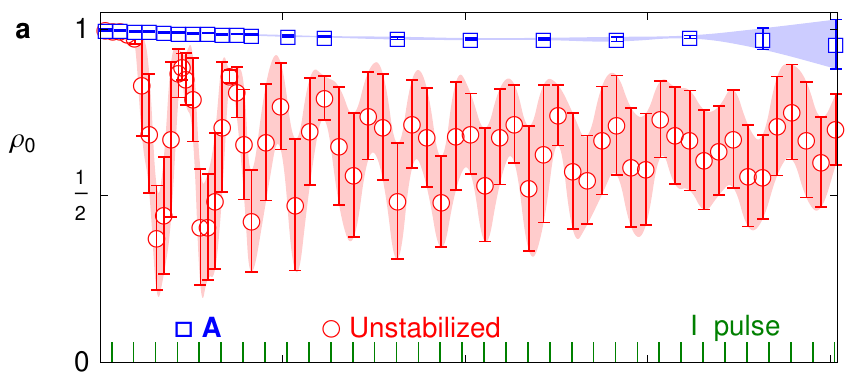}\\
	 		\includegraphics[width=3.4in]{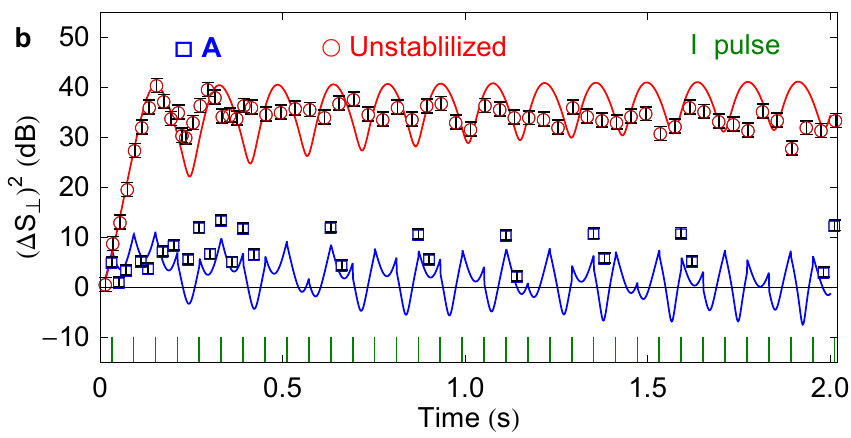}	
	\end{minipage}	
	
	\caption{{\bf Long timescale stabilization by microwave pulses.} (a) The stabilization dynamic population $\rho_0$, the letter A corresponds to the location in stability diagram \figref[c]{fig:StabilityMapuWave}. The shaded region are derived from standard deviation to guide the eyes. (b) The uncertainty of transverse magnetization $\Delta \Sperp_\bot$. Solid line is quantum simulation.}
    \label{fig:StablizationuWave}	
\end{figure}

Although the effectiveness of the control in presence of added noise is not the focus of this investigation, we note that at least for uncorrelated atom loss, the stabilization method is still effective.  This is shown in \figref{fig:StablizationuWave} where we observe  long term stabilization for the most stable parameters (A).  With these parameters the $\rho_0=1$ state is stabilized for nearly 2~s before any appreciable deviation is observed.  In this time period the unstabilized condensate  undergoes thirteen oscillations with apparent damping towards the ground state population.  Furthermore the stabilization is maintained even though $75\%$ of the initial number of atoms have been lost due to the finite trap lifetime of $1.4~\mathrm{s}$.

Compared to previous proposals for stabilizing dynamics in the double well system \cite{PhysRevA.80.053613} or the spin-1 condensate \cite{Zhang10} based on periodic reversals of the \emph{sign} of $q$ (or $\lambda$), our method is based on changing the \emph{magnitude} $q$. The effect of the periodic microwave pulses can be approximated by a time-averaged Hamiltonian with an effective quadratic Zeeman energy, $q_\mathrm{eff} = q + \hbar \Delta\theta/\tau$. It is an interesting question, however,  whether or not our observed stability is explained solely by this effect.  For both $q_\mathrm{eff} > 2 |c|$ and $q_\mathrm{eff} < 0$ there is no longer a hyperbolic fixed point centered on the pure $m_f=0$ state but rather an elliptical fixed point, and hence wherever these conditions are met, the time-averaged system will be inherently stable.  This defines a `robust' stability region that is shown as dashed red lines in \figref[c]{fig:StabilityMapuWave}. Although this region agrees asymptotically with the linear stability analysis (solid red lines) for shorter pulse periods, the robust region is much smaller. Except for the determination the stability map, all of the measurements presented are outside of the robust region, which shows that the time-averaged Hamiltonian is insufficient to describe the results.

In summary, we have demonstrated dynamical stabilization of a quantum many-body system consisting of an interacting spin-1 Bose condesate, and we have begun to explore the range of parameters for which it can be stabilized.
Although the stabilization is demonstrated with a microcondensate for which the spatial dynamics are factored out, these methods should be applicable to the control of the coupled spin/spatial dynamics that lead to domain formation in larger condensates.  In future investigations, it would be interesting to explore this area as well as the application of these concepts to finite temperature spin systems.

\section{Methods}
\small
The experiment begins with a condensate containing $N=4.5 \times 10^4$ atoms initialized in the $|f=1,m_f=0\rangle$ hyperfine state held in a high magnetic field ($2$~G). To initiate spin dynamics, the condensate is rapidly quenched below the quantum critical point by lowering the magnetic field to 220~mG. In order to stabilize the dynamics, the $\{\Sperp_\bot,\Q_\bot\}$ quadrature phase is periodically rotated by an angle $\Delta \theta$ with a period $\tau$.
The rotation is implemented using $2\pi$ Rabi pulses on the $|f=1,m_f=0\rangle \leftrightarrow |f=2,m_f=0\rangle$ microwave clock transition that effectively shift the phase of the $|f=1,m_f=0\rangle$ spinor component by an amount $\Delta\theta_0=\pi \left(1+ \Delta / \sqrt{1+\Delta^2} \right)$ where $\Delta = \delta/\Omega$ is the detuning normalized to the on-resonance Rabi rate \cite{Hamley12}. This pulse effects a quadrature rotation $\Delta\theta = -\Delta\theta_0$. The quadrature rotation is varied from $-\pi/2$ to $-3\pi/2$.
Finally, the spin populations of the condensate are measured.  This is executed by releasing the trap and allowing the atoms to freely expand in a Stern-Gerlach magnetic field gradient to separate the $m_f$ spin components.  The atoms are probed for $400~\mu\mathrm{s}$ by three pairs of orthogonal laser beams, and the fluorescence signal is collected by a CCD camera.
In order to measure $\Delta \Sperp_\bot$ an RF rotation is used to rotate it into $S_z$ which is measured by counting the difference in the imaged $m_f=1$ and $m_f=-1$ atoms.
The background magnetic field $B=220(10)$ mG is determined by RF and microwave spectroscopy while the spinor dynamical rate 
is determined by experimental coherent oscillation prepared near the ground state ($c=-2\pi \times 7.2(2)~\mathrm{Hz}$).

\section{Acknowledgements}

We would like to thank Paul Goldbart and Rafael Hipolito for assistance with the manuscript, as well as Carlos S\'a de Melo for helpful discussions.  We acknowledge support from the NSF.

\bibliography{QPref}
\bibliographystyle{nature}

\end{document}


\title{Dynamic Stabilization of a Quantum Many-body System:  Supplementary Information}
\author{T.M. Hoang, C.S. Gerving, B.J. Land, M. Anquez, C.D. Hamley,  and M.S. Chapman}
\affiliation{School of Physics, Georgia Institute of Technology,
  Atlanta, GA 30332-0430}

\maketitle

In this Supplementary Information, 
we derive the stability regions shown in the main paper for pulsed quadrature phase shifts applied periodically during the evolution.  The derivation follows standard methods of stability analysis for step-wise evolution used for example in optical resonator stability analysis. We compare this method with an analysis based on the time-average of the Hamiltonian that provides more restrictive condition in disagreement with the experiment.

\section{Model of the System}

The Hamiltonian of the system in the single mode approximation is given by
\begin{equation}
	\mathcal{H} = \lambda \hat{\Slab}^2 + p \hat{\Slab}_z + \frac{q}{2} \hat{\Qlab}_{zz} \nonumber
\end{equation}
where $\hat{\Slab}^2=\hat{\Slab}_x^2+\hat{\Slab}_y^2+\hat{\Slab}_z^2$ is the total spin operator with $\hat{\Slab}_z$ its projection along $B$, $\hat{\Qlab}_{zz}$ is a moment of the spin-1 nematic or quadrupole tensor, $\lambda \propto a_2-a_0$ is the spin interaction strength integrated over the condensate, $a_F$ is the total spin $F$ $s$-wave scattering length, $p \propto B$ and $q \propto B^2$ are the per atom linear and quadratic Zeeman energies, respectively.  
The spin-1 system has SU(3) symmetry where the Hamiltonian and operators can be represented by an su(3) Lie algebra for which we choose a dipole-quadrupole basis.  Definitions of the operators and a commutator table are given in the Appendix.  $\Slab_z$ commutes with the Hamiltonian and has trivial contribution to the dynamics (Larmor precession). Going to a rotating frame, we take
\begin{eqnarray}
	\mathcal{H}_0 &=& p \hat{\Slab}_z \nonumber\\
					\mathcal{H}_{I} &=& \hat{U}^\dagger \left(\lambda \hat{\Slab}^2 + \frac{q}{2}\hat{\Qlab}_{zz}  \right) \hat{U} \notag\\
 						&=& \lambda \hat{S}^2 + \frac{q}{2}\hat{Q}_{zz} \nonumber 
\end{eqnarray}
where $\hat{U} = \exp(-\frac{\imath\mathcal{\hat{H}}_0 t}{\hbar}) = \exp(-\frac{\imath p \hat{\mathcal{S}}_z t}{\hbar})$.  
We define operators in rotating frame $\hat{A} = \hat{U}^\dag \hat{\mathcal{A}} \hat{U}$ , which are given explicitly in the Appendix.  
$\hat{S}^2$ and $\hat{Q}_{zz}$ are invariants under this unitary transformation.  

Using these operators, we calculate the Heisenberg equations of motion $\dot{\hat{A}} = \frac{\imath}{\hbar}\left[ \mathcal{H}_I, \hat{A} \right] $ which yields,

{\allowdisplaybreaks
\begin{eqnarray}
	\dot{\hat{S}}_x &=& - \frac{q}{\hbar} \hat{Q}_{yz} \notag\\
	\dot{\hat{S}}_y &=& \frac{q}{\hbar} \hat{Q}_{xz} \notag\\
	\dot{\hat{S}}_z &=& 0 \notag\\
	\dot{\hat{Q}}_{yz} &=& -\frac{\lambda}{\hbar} (\lbrace \hat{Q}_{0+},\hat{S}_x \rbrace + \lbrace \hat{Q}_{xy},\hat{S}_y \rbrace  \notag\\ 
	&&- \lbrace \hat{Q}_{xz},\hat{S}_z \rbrace)+ \frac{q}{\hbar} \hat{S}_{x} \notag \\
	\dot{\hat{\Qrot}}_{xz}  &=& -\frac{\lambda}{\hbar} (\lbrace \hat{\Qrot}_{0-},\hat{\Srot}_y \rbrace - \lbrace \hat{\Qrot}_{xy},\hat{\Srot}_x \rbrace \notag \\ 		
	&&+ \lbrace \hat{\Qrot}_{yz},\hat{\Srot}_z \rbrace)  - \frac{q}{\hbar} \hat{\Srot}_{y} \notag \\
\dot{\hat{\Qrot}}_{0+} &=& \frac{2 \lambda}{\hbar} (2 \lbrace \hat{\Qrot}_{yz},\hat{\Srot}_x \rbrace - \lbrace \hat{\Qrot}_{xz},\hat{\Srot}_y \rbrace- \lbrace \hat{\Qrot}_{xy},\hat{\Srot}_z \rbrace) \notag\\
		\dot{\hat{\Qrot}}_{0-} &=& \frac{2\lambda}{\hbar} (2 \lbrace \hat{\Qrot}_{xz},\hat{\Srot}_y \rbrace - \lbrace \hat{\Qrot}_{yz},\hat{\Srot}_x \rbrace - \lbrace \hat{\Qrot}_{xy},\hat{\Srot}_z \rbrace) \notag\\
	\dot{\hat{\Qrot}}_{xy} &=& \frac{\lambda}{\hbar} (\lbrace \hat{\Qrot}_{0+}+\hat{\Qrot}_{0-},\hat{\Srot}_z \rbrace
	 -\lbrace \hat{\Qrot}_{xz},\hat{\Srot}_x \rbrace \nonumber\\
	 & &+ \lbrace \hat{\Qrot}_{yz},\hat{\Srot}_y \rbrace)  
	 \label{QEquationMotion}	 
\end{eqnarray}
}
Where the anticommutator  $\lbrace \hat{A}, \hat{B} \rbrace = \hat{A}\hat{B}+\hat{B}\hat{A}$

\section{Stability Analysis}
\subsection{Stabilization Dynamics}
We are primarily concerned with the stabilization about our initial state where $N_0 = N$.  In order to linearize the equations of motion we expand the operators about the expectation values of this initial state, $\hat{A} = \langle \hat{A} \rangle + \delta\hat{A} $. The only nonzero expectation values are
\begin{eqnarray}
	\langle \hat{\Qrot}_{0+} \rangle
	&=& - 2 N \notag \\
	\langle \hat{\Qrot}_{0-} \rangle
	&=&  2 N. \notag 	
\end{eqnarray}
Keeping the terms linear in $\delta\hat{A}$, and eliminating the higher order terms of $\delta\hat{A}$ and the products of  $\delta\hat{A} \cdot \delta\hat{B}$. The dynamical equations \eqnref{QEquationMotion} become
\begin{eqnarray}
	\delta\dot{\hat{\Srot}}_x &=& - \frac{q}{\hbar}\delta \hat{\Qrot}_{yz} \notag\\
	\delta\dot{\hat{\Srot}}_y &=& \frac{q}{\hbar}\delta \hat{\Qrot}_{xz} \notag\\
	\delta\dot{\hat{\Qrot}}_{yz} &=& \frac{4 N\lambda}{\hbar} \delta\hat{\Srot}_x  + \frac{q}{\hbar} \delta\hat{\Srot}_{x} \notag \\
	\delta\dot{\hat{\Qrot}}_{xz}  &=& -\frac{4N\lambda}{\hbar} \delta\hat{\Srot}_y  - \frac{q}{\hbar} \delta\hat{\Srot}_{y} \notag \\
\delta\dot{\hat{\Qrot}}_{0+} &=& 0 \notag\\
		\delta\dot{\hat{\Qrot}}_{0-}&=&0 \notag\\
	\delta\dot{\hat{\Qrot}}_{xy} &=&0 \notag
\end{eqnarray}
These equations describe the quantum dynamics in the neighborhood of the pole at which squeezing happens. In order to determine the stability condition we make a mean-field approximation, where we replace the operator $\delta\hat{A}$ with its expectation value $\delta A$.  Also since $\langle \hat{A}\rangle=0$ for all the dynamical operators, we will drop the $\delta$ notation of the expansion. 
By defining
\begin{eqnarray}
	\Srot_x &=& \Srot_\bot \cos(\theta_L)_{t=0} \nonumber \\
	\Srot_y &=& - \Srot_\bot \sin(\theta_L)_{t=0} \nonumber \\
	\Qrot_{yz} &=& - \Qrot_\bot \cos(\theta_L)_{t=0} \nonumber \\
	\Qrot_{xz} &=& - \Qrot_\bot \sin(\theta_L)_{t=0} \nonumber
\end{eqnarray}
The system of equations simplify into
\begin{eqnarray}
	\dot{{\Srot}}_\bot  &=& \tilde{q} \Qrot_{\bot} \nonumber\\
	\dot{\Qrot}_{\bot}  &=& -\left(2 \tilde{c}  + \tilde{q}\right) {\Srot}_{\bot}
	\label{SbotQbotEqn} 
\end{eqnarray}
with spinor dynamical rate $c =2 N\lambda$, and  $\tilde{c} = c/\hbar$, and $\tilde{q} = q/\hbar$ are angular frequencies. 

\subsection{Stabilization Condition}
\begin{figure}[t!]
	\begin{center}
	 	\includegraphics[scale=1]{StabilityMapuWaveLegend.pdf}\\		
		\includegraphics[scale=1]{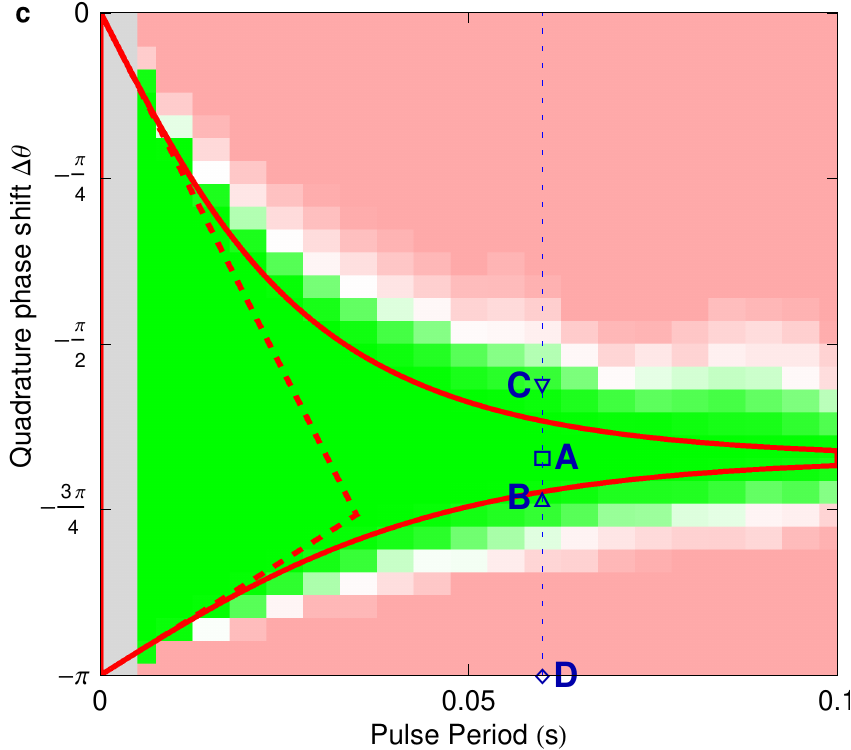} 		 			 	
	\end{center}
    \caption{\textbf{Stability mapping.}  Simulation values of $\rho_0$ at the time of the maximum spin mixing are used to map the stability region.  This is compared to the linear stability analysis (solid line) and the time averaged $q_\mathrm{eff}$ stability analysis (dashed line).}
\label{StabilityMap}
\end{figure}

\eqnref{SbotQbotEqn} can be written in a matrix form
\begin{equation}
	\left(
		\begin{array}{c}
			\dot{\Srot}_\bot \\
			\dot{\Qrot}_{\bot}
		\end{array}
	\right)
	= \left(
		\begin{array}{cc}
			0 & \tilde{q} \\ \\
			-\left( 2\tilde{c} + \tilde{q} \right) & 0
		\end{array}
	\right)
	\left(
		\begin{array}{c}
			\Srot_\bot \\
			\Qrot_{\bot}
		\end{array}
	\right) \nonumber
\end{equation}
which is more amenable to the stability analysis to follow.
Defining this matrix as $\mathbf{m}$, the time evolution is given by its exponentiation.
 The quadrature phase shift used for stabilization is given by the operator $\exp\left( \imath \Delta\theta \hat{\Qrot}_{zz} \right)$, which is just a plane rotation in $\lbrace \Srot_\bot, \Qrot_{\bot} \rbrace$ by an angle $\Delta\theta$.  The full dynamics from one pulse to another including the quadrature phase shift is given by
\begin{equation}
\mathbf{M} = \mathbf{R}[\Delta\theta] \cdot \exp [\tau \mathbf{m}] \nonumber
\end{equation}
where $\tau$ is the period between pulses, $\mathbf{R}$ is a 2 dimensional rotation matrix, and $\Delta\theta$ is the amount of the quadrature phase shift.
From this point the stability condition is exactly the same as an optical resonator using ray tracing techniques.  The determinant of the one pass evolution matrix must meet the condition $|\mathrm{Tr}[\mathbf{M}]|<2$. Evaluating this condition gives the inequality,
\begin{equation}
	2\left| \cos \Delta\theta  \cosh \Gamma \tau+\frac{\tilde{c}+\tilde{q}}{\Gamma} \sin \Delta\theta \sinh \Gamma \tau \right|< 2\nonumber
\label{StabilityInequality}
\end{equation}
with $\Gamma=\sqrt{\tilde{q}(2|\tilde{c}|-\tilde{q})}$.  This inequality is used to mark the boundaries of the analytic stability region, which compares favorably to simulations \figref{StabilityMap}.

\subsection{Effective quadratic Zeeman analysis}
Another way to analyze the stability is through the use of a time-averaged, effective quadratic Zeeman.  For $c<0$, it is known that the system does not evolve from the $m_f=0$ state if the quadratic Zeeman energy is greater than $2|c|$ or less than zero.  We note that the quadratic Zeeman portion of $\mathbf{M}$ is just a two dimensional rotation matrix rotating the quadratures by an angle of $q\,\tau/\hbar$.  The microwave pulse generates an instantaneous quadrature rotation of $\Delta\theta$.  Therefore over one period of the periodic microwave pulse sequence we can calculate an effective quadratic Zeeman, $q_\mathrm{eff} = q + \hbar \Delta\theta/\tau$.  
The phase space is cyclic with a periodicity of $\pi$ in quadrature phase.  So an instantaneous phase shift of $\Delta\theta$ is equivalent to a phase shift of $\Delta\theta-\pi$, which is a negative contribution to $q_\mathrm{eff}$.  
Because of this ambiguity of the phase shift direction, $q_\mathrm{eff}$ is double valued everywhere.  Both values are needed for the time-averaged analysis of the stability.
For both $q_\mathrm{eff} > 2 |c|$ and $q_\mathrm{eff} < 0$ there is no longer a hyperbolic fixed point centered on the $m_f=0$ state.  Wherever these conditions are met for both of the $q_\mathrm{eff}$ values, then the system is trivially stable and define a `robust' stability region.  This region is bounded by the dashed lines in the stability diagram.  The dynamical solutions asymptote to these lines for short time between pulses.

\section{Classical phase space}

Within the single mode approximation (in which the spin states share a common spatial wavefunction), the mean-field order parameter is represented by a complex vector $\vec{\zeta}=(\zeta_1,\zeta_0,\zeta_{-1})^T$ where $\zeta_i$ represents the amplitude and phase of the classical field for the mode associated with the $m_f$ state given in the index and $|\vec{\zeta}|^2=1$.  

A convenient parameterization of the order parameter is given by
\begin{eqnarray}
	\zeta_1 &=& \sqrt{\frac{1-\rho_0+m}{2}} e^{\imath (\theta+\theta_L)} \nonumber \\
	\zeta_0 &=& \sqrt{\rho_0} \nonumber \\
	\zeta_{-1} &=& \sqrt{\frac{1-\rho_0-m}{2}} e^{\imath (\theta-\theta_L)} \nonumber
\label{orderparam}
\end{eqnarray}
where $\theta_L = (\theta_1-\theta_{-1})/2$ is the Larmor precession phase, $\theta = (\theta_1+\theta_{-1}-2\theta_0)/2$ is the quadrature phase,
and $m=(N_1-N_{-1})/N = \rho_{1}-\rho_{-1}$ is the fractional magnetization, which is a constant of the motion.

Also using this parameterization we define the mean-field expectation values of $\Srot_x$, $\Srot_y$, $\Qrot_{yz}$, and $\Qrot_{xz}$ in order to derive the spin sphere representation used in the main paper.  The corresponding mean field expectation values when $m=0$ are given by
\begin{eqnarray}
	 \Srot_x  &=&  \langle \vec{\zeta} \mid \mathbf{S}_x \mid \vec{\zeta} \rangle = 2 \sqrt{\rho_0 (1- \rho_0)} \cos \theta \cos \theta_L \nonumber \\
	 \Qrot_{yz}  &=& -2 \sqrt{\rho_0 (1- \rho_0)} \sin\theta \cos \theta_L \nonumber \\
	\Srot_y &=& -2 \sqrt{\rho_0 (1- \rho_0)} \cos \theta \sin\theta_L \nonumber \\
	\Qrot_{xz}  &=& -2 \sqrt{\rho_0 (1- \rho_0)} \sin\theta \sin\theta_L \nonumber 
\label{MFQuads}
\end{eqnarray}

Since $\Srot_\perp^2=\Srot_x^2+\Srot_y^2$ and $\Qrot_\perp^2=\Qrot_{xz}^2+\Qrot_{yz}^2$, we define $\Srot_\perp = 2 \sqrt{\rho_0 (1- \rho_0)} \cos \theta$ and $\Qrot_\perp = 2 \sqrt{\rho_0 (1- \rho_0)} \sin \theta$.  
Defining $x = 2\rho_0-1$, then 
\begin{eqnarray}
	\Srot^2_\bot+\Qrot^2_\bot + x^2 = 1. \nonumber
\end{eqnarray}  
$\Srot_\perp$, $\Qrot_\perp$, and $x$ have spin Poisson brackets and thus define a classical spin representation.  This representation is shown as a sphere in Fig.~1 of the main paper.

\onecolumngrid
\pagebreak
\appendix*
\section{Spin and Quadrupole Operators}
\label{SpinQuadrupole}
\begin{table*}[h!!!]
\caption{Spin-1 dipole operators.  Expectation values of these operators are components of the angular momentum vector.  Matrices in spherical polar basis $|f,m_f\rangle$.}
\begin{center}
\begin{tabular}{l@{}c@{\hspace{0.5in}}cl}
$\mathbf{S}_x = \frac{1}{\sqrt{2}}$ & $\left(
\begin{array}{ccc}
	0 & 1 & 0 \\
	1 & 0 & 1 \\
	0 & 1 & 0
\end{array}
\right)$
& &
$\hat{\Slab}_x = \frac{1}{\sqrt{2}} \left( \hat{a}_{1}^{\dag} \hat{a}_{0}
	+\hat{a}_{0}^{\dag} \hat{a}_{-1} + \hat{a}_{0}^{\dag} \hat{a}_{1}
	+ \hat{a}_{-1}^{\dag} \hat{a}_{0} \right)$
\\
$\mathbf{S}_y = \frac{\imath}{\sqrt{2}}$ & $\left(
\begin{array}{ccc}
	0 & -1 & 0 \\
	1 & 0 & -1 \\
	0 & 1 & 0
\end{array}
\right)$
& &
$\hat{\Slab}_y = \frac{\imath}{\sqrt{2}} \left( - \hat{a}_{1}^{\dag} \hat{a}_{0}
	- \hat{a}_{0}^{\dag} \hat{a}_{-1} + \hat{a}_{0}^{\dag} \hat{a}_{1}
	+ \hat{a}_{-1}^{\dag} \hat{a}_{0} \right)$
\\
$\mathbf{S}_z =$ & $\left(
\begin{array}{ccc}
	1 & 0 & 0 \\
	0 & 0 & 0 \\
	0 & 0 & -1
\end{array}
\right)$
& &
$\hat{\Slab}_z = \left( \hat{a}_{1}^{\dag} \hat{a}_{1} - \hat{a}_{-1}^{\dag} \hat{a}_{-1} \right)$
\end{tabular}
\end{center}
\label{Dipole}
\vspace{0.1in}

\caption{The spin-1 quadrapole operators.  Expectation values of these operators are moments of the symmetric traceless quadrapole tensor.  Matrices in spherical polar basis $|f,m_f\rangle$.}
\begin{center}
\begin{tabular}{l@{}c@{}c@{\hspace{0.5in}}cl}
$\mathbf{Q}_{yz} =$ & $\frac{\imath}{\sqrt{2}}$ & $\left(
\begin{array}{ccc}
	0 & -1 & 0 \\
	1 & 0 & 1 \\
	0 & -1 & 0
\end{array}
\right)$
& &
$\hat{\Qlab}_{yz} = \frac{\imath}{\sqrt{2}} \left( - \hat{a}_{1}^{\dag} \hat{a}_{0}
	+ \hat{a}_{0}^{\dag} \hat{a}_{-1} + \hat{a}_{0}^{\dag} \hat{a}_{1}
	- \hat{a}_{-1}^{\dag} \hat{a}_{0} \right)$
\\
$\mathbf{Q}_{xz} =$ & $\frac{1}{\sqrt{2}}$ & $\left(
\begin{array}{ccc}
	0 & 1 & 0 \\
	1 & 0 & -1 \\
	0 & -1 & 0
\end{array}
\right)$
& &
$\hat{\Qlab}_{xz} = \frac{1}{\sqrt{2}} \left( \hat{a}_{1}^{\dag} \hat{a}_{0}
	- \hat{a}_{0}^{\dag} \hat{a}_{-1} + \hat{a}_{0}^{\dag} \hat{a}_{1}
	- \hat{a}_{-1}^{\dag} \hat{a}_{0} \right)$
\\
$\mathbf{Q}_{xy} =$ & $\imath$ & $\left(
\begin{array}{ccc}
	0 & 0 & -1 \\
	0 & 0 & 0 \\
	1 & 0 & 0
\end{array}
\right)$
& &
$\hat{\Qlab}_{xy} = \imath \left( - \hat{a}_{1}^{\dag} \hat{a}_{-1}
	+ \hat{a}_{-1}^{\dag} \hat{a}_{1} \right)$
\\
$\mathbf{Q}_{xx} =$ & & $\left(
\begin{array}{ccc}
	-\frac{1}{3} & 0 & 1 \\
	0 & \frac{2}{3} & 0 \\
	1 & 0 & -\frac{1}{3}
\end{array}
\right)$
& &
$\hat{\Qlab}_{xx} = -\frac{1}{3} \hat{a}_{+1}^\dag \hat{a}_{+1}
	+\frac{2}{3} \hat{a}_0^\dag \hat{a}_0 -\frac{1}{3} \hat{a}_{-1}^\dag a_{-1}
	+ \hat{a}_{+1}^{\dag} \hat{a}_{-1} + \hat{a}_{-1}^{\dag} \hat{a}_{+1}$
\\
$\mathbf{Q}_{yy} =$ & & $\left(
\begin{array}{ccc}
	-\frac{1}{3} & 0 & -1 \\
	0 & \frac{2}{3} & 0 \\
	-1 & 0 & -\frac{1}{3}
\end{array}
\right)$
& &
$\hat{\Qlab}_{yy} = -\frac{1}{3} \hat{a}_{+1}^\dag \hat{a}_{+1}
	+\frac{2}{3} \hat{a}_0^\dag \hat{a}_0 -\frac{1}{3} \hat{a}_{-1}^\dag \hat{a}_{-1}
	- \hat{a}_{+1}^{\dag} \hat{a}_{-1} - \hat{a}_{-1}^{\dag} \hat{a}_{+1}$
\\
$\mathbf{Q}_{zz} =$ & & $\left(
\begin{array}{ccc}
	\frac{2}{3} & ~0 & 0 \\
	0 & -\frac{4}{3} & 0 \\
	0 & ~0 & \frac{2}{3}
\end{array}
\right)$
& &
$\hat{\Qlab}_{zz} = \frac{2}{3} \hat{a}_{+1}^\dag \hat{a}_{+1}
	-\frac{4}{3} \hat{a}_0^\dag \hat{a}_0 +\frac{2}{3} \hat{a}_{-1}^\dag \hat{a}_{-1}$
\end{tabular}
\end{center}
\label{Quadrapole}
\vspace{0.1in}

\begin{center}
\caption{Commutators of the dipole-quadrupole basis.  
The commutator table is applicable to both the quantum operators and the matrix form.  Furthermore the unitary transformation used to go to a rotating frame does not change its structure.}
\vspace{0.1in}
\begin{tabular}{|c||c|c|c|c|c|c|c|c|}
	\hline
	$\left[ \downarrow,\rightarrow \right]$ & $S_y$ & $S_z$ & $Q_{yz}$ & $Q_{xz}$ & $Q_{xy}$ & $Q_{xx}$ & $Q_{yy}$ & $Q_{zz}$ \\
	\hline \hline
	$S_x$ & $iS_z$ & $-iS_y$ & $i(Q_{zz}-Q_{yy})$ & $-iQ_{xy}$ & $iQ_{xz}$ & 0 & $2iQ_{yz}$ & $-2iQ_{yz}$ \\
	\hline
	$S_y$ & & $iS_x$ & $iQ_{xy}$ & $i(Q_{xx}-Q_{zz})$ & $-iQ_{yz}$ & $-2iQ_{xz}$ & 0 & $2iQ_{xz}$ \\
	\hline
	$S_z$ & & & $-iQ_{xz}$ & $iQ_{yz}$ & $i(Q_{yy}-Q_{xx})$ & $2iQ_{xy}$ & $-2iQ_{xy}$ & 0 \\
	\hline
	$Q_{yz}$ & & & & $-iS_z$ & $iS_y$ & 0 & $-2iS_x$ & $2iS_x$ \\
	\hline
	$Q_{xz}$ & & & & & $-iS_x$ & $2iS_y$ & 0 & $-2iS_y$ \\
	\hline
	$Q_{xy}$ & & & & & & $-2iS_z$ & $2iS_z$ & 0 \\
	\hline
	$Q_{xx}$ & & & & & & & 0 & 0 \\
	\hline
	$Q_{yy}$ & & & & & & & & 0 \\
	\hline
\end{tabular}
\end{center}
\label{Commutators}
\end{table*}

\begin{table*}[h!!!]
\caption{Operators in rotating frame using the Baker-Hausdorff lemma.}
\begin{eqnarray}
	\hat{\Srot}_x &=& \hat{\Slab}_x \cos(\frac{pt}{\hbar}) - \hat{\Slab}_y \sin(\frac{pt}{\hbar}) \notag \\
	\hat{\Srot}_y &=& \hat{\Slab}_x \sin(\frac{pt}{\hbar}) + \hat{\Slab}_y \cos(\frac{pt}{\hbar}) \notag \\
	\hat{\Qrot}_{yz} &=& \hat{\Qlab}_{xz} \sin(\frac{pt}{\hbar}) + \hat{\Qlab}_{yz} \cos(\frac{pt}{\hbar}) \notag \\	
	\hat{\Qrot}_{xz} &=& \hat{\Qlab}_{xz} \cos(\frac{pt}{\hbar}) - \hat{\Qlab}_{yz} \sin(\frac{pt}{\hbar}) \notag \\		
	\hat{\Qrot}_{0+} &=& \hat{\Qrot}_{zz}-\hat{\Qrot}_{yy} = \frac{1}{2}(\hat{\Qlab}_{0+}-\hat{\Qlab}_{0-})  \notag\\ 
	&&+\frac{1}{2} (\hat{\Qlab}_{0+}+\hat{\Qlab}_{0-}) \cos(\frac{2pt}{\hbar}) - \hat{\Qlab}_{xy}\sin(\frac{2pt}{\hbar}) \notag \\
	\hat{\Qrot}_{0-}&=&	\hat{\Qrot}_{xx}-\hat{\Qrot}_{zz} = -\frac{1}{2}(\hat{\Qlab}_{0+}-\hat{\Qlab}_{0-})  \notag\\ 
	&&+\frac{1}{2} (\hat{\Qlab}_{0+}+\hat{\Qlab}_{0-}) \cos(\frac{2pt}{\hbar})- \hat{\Qlab}_{xy}\sin(\frac{2pt}{\hbar}) \notag \\			
	\hat{\Qrot}_{xy} &=& \hat{\Qlab}_{xy} \cos(\frac{2pt}{\hbar}) + \frac{1}{2} (\hat{\Qlab}_{0+}+\hat{\Qlab}_{0-}) \sin(\frac{2pt}{\hbar}) \notag 		
\end{eqnarray}
\end{table*}

\twocolumngrid
\bibliography{QPref}
\bibliographystyle{unsrt}